\documentclass[]{aastex631}
\usepackage{epstopdf}
\usepackage{tikz}
\usepackage{amsmath}
\usepackage{bm}

\newcommand\aastex{AAS\TeX}

\received{xx}
\revised{xx}
\accepted{xx}
\submitjournal{xxx}

\shorttitle{\aastex\ sample article}
\shortauthors{Cai}
\begin{document}

\title{Spontaneous Generated Convective Anticyclones in Low Latitude --- A Model for the Great Red Spot}

\correspondingauthor{Tao Cai}
\email{tcai@must.edu.mo}

\author[0000-0003-3431-8570]{Tao Cai}
\affil{State Key Laboratory of Lunar and Planetary Sciences, Macau University of Science and Technology, Macau, People's Republic of China}

\author[0000-0002-6428-1812]{Kwing L. Chan}
\affil{State Key Laboratory of Lunar and Planetary Sciences, Macau University of Science and Technology, Macau, People's Republic of China}
\affil{Faculty of Information and Technology, Macau University of Science and Technology, Macau, People's Republic of China}

\author[0000-0003-2512-5670]{Kim-Chiu Chow}
\affil{State Key Laboratory of Lunar and Planetary Sciences, Macau University of Science and Technology, Macau, People's Republic of China}



\begin{abstract}
The Great Red Spot at about latitude $22^{\circ}S$ of Jupiter has been observed for hundreds of years, yet the driving mechanism on the formation of this giant anticyclone still remains unclear. Two scenarios were proposed to explain its formation. One is a shallow model suggesting that it might be a weather feature formed through a merging process of small shallow storms generated by moist convection, while the other is a deep model suggesting that it might be a deeply rooted anticyclone powered by the internal heat of Jupiter. In this work, we present numerical simulations showing that the Great Red Spot could be naturally generated in a deep rotating turbulent flow and survive for a long time, when the convective Rossby number is smaller than a certain critical value. From this critical value, we predict that the Great Red Spot extends at least about 500 kilometers deep into the Jovian atmosphere. Our results demonstrate that the Great Red Spot is likely to be a feature deep-seated in the Jovian atmosphere.
\end{abstract}



\section{Introduction}
Vortices are ubiquitous in gas giant planets. In Jupiter, more than 500 vortices have been identified from the sequence of 70-day Cassini images with latitude spanning from $80^{\circ}S$ to $80^{\circ}N$ \citep{li2004life}. Recent observation from the Juno spacecraft has detected multiple closely-packed cyclones in the northern and southern polar regions \citep{adriani2018clusters,tabataba2020long}. Anticyclones of sizes larger than 1000km were also observed in the polar regions of Jupiter, among which most were found to form a dipole configuration with a cyclone \citep{adriani2020two}. The Great Red Spot (GRS), is probably the most prominent large-scale vortex in Jupiter, which is an anticyclone centered at about $22^{\circ}S$  and has been observed for hundreds of years. The origin of GRS, however, remains a puzzle.

To understand the origin and existence of large-scale vortices (LSVs) in Jupiter, two scenarios of flow dynamics have been proposed based on different assumptions on the depth of the Jovian atmosphere.  The first is a shallow model in which jets \citep{liu2010mechanisms,lian2010generation} and vortices \citep{zhang2014atmospheric,o2015polar} are driven by the energy generated from the latent heat of moist convection. The second is a deep model in which columnar jets \citep{busse1976simple,christensen2001zonal,heimpel2007turbulent,cai2012three,aurnou2015rotating} and vortices \citep{chan2013numerical,heimpel2016simulation,yadav2020deepa,cai2021deep} are maintained by the energy deep from the interior of Jupiter.

While the depth of the Jovian atmosphere is uncertain, recent measurement on gravity field by Juno indicates that the zonal jets could possibly extend thousands of kilometers deep into the atmosphere \citep{kaspi2018jupiter,guillot2018suppression}. The heating of Jupiter's upper atmosphere also reveals that the GRS should be heated from below, suggesting that it might have a deep origin \citep{o2016heating}. Although there are some evidences favoring the deep model, it is fair to say that the possibility of the shallow scenario is not entirely excluded \citep{kong2018origin}.

Even if one assumes that the atmosphere is deep, it is natural to ask whether LSVs, such as GRS could be formed. Previous simulations on rotating compressible convection \citep{chan2007rotating,kapyla2011starspots} revealed that LSVs could be generated in a rapidly rotating turbulent flow. Later, the dynamics of LSVs has been intensively studied in a rapidly rotating Rayleigh-B\'enard convection \citep{guervilly2014large,rubio2014upscale,favier2014inverse}. Despite the considerable progress achieved in these simulations, there are some limitations in these studies. First, the heat fluxes used in these simulations \citep{chan2007rotating,kapyla2011starspots} are larger than Jupiter's internal heat flux by several magnitudes. Second, in previous simulations on rapidly rotating convection only shear \citep{novi2019rapidly,currie2020generation} or large-scale cyclones \citep{chan2013numerical} can be found in low latitudes. The present study is motivated by the following questions. Could LSVs be generated in deep convective flow with a flux at or close to Jupiter's internal heat? Could large-scale anticyclones, such as GRS, be generated in a rapidly rotating turbulent flow at low latitudes? Results of some numerical simulations will be discussed in the following sections to address these questions.

\section{Result}
\subsection{How to generate LSVs at Jupiter-like small flux?}
We first consider the simulations of rotating turbulent convection in a {\it f}-plane at high latitudes. The computational domain is a Cartesian box with a constant flux fed at the bottom. Large eddy simulations were performed with a semi-implicit mixed finite-difference spectral code \citep{cai2016semi}, which accounts for the effects of density stratification, compressibility, and subgrid scale turbulence (see Appendix \ref{appendixA}). We use dimensionless variables in the code, with all the variables normalized by the values of height, pressure, density, and temperature at the top of the box. For example, the velocity is normalized by $(p_{top}/\rho_{top})^{1/2}$, and the flux is normalized by $\rho_{top}(p_{top}/\rho_{top})^{3/2}$, where $p_{top}$ and $\rho_{top}$ are the pressure and density at the top of box, respectively. The pressure and density at Jupiter's surface (defined at the location with 1 bar level pressure) are $10^5 {\rm Pa}$ and $0.167 {\rm kg/m^3}$ respectively. The internal heat flux of Jupiter is estimated to be $7.48 {\rm Wm^{-2}}$ \citep{li2018less}. For Jupiter, the dimensionless total flux $F_{tot_J}$ is about $1\times 10^{-7}$, which is mostly carried by convection in the convection zone. Due to the low Mach number and long period of integration time, simulation at small flux is still a challenging problem in stellar or planetary convection \citep{kupka2017modelling}. So far, only a few attempts were made on the simulations of solar-like low flux convection \citep{hotta2017solar,kapyla2019overshooting}. Simulations on rotating convection at low fluxes have not yet been reported.

The simulation box has an aspect ratio of $\Gamma=4$ (lateral dimension to height), and a resolution of $512^2\times 101$ grid points. By using this resolution, the simulation marginally shows the Kolmogrov inertial range and the small scale effects are approximated by a simple small scale turbulence model (see Appendix \ref{appendixD} and Fig.~\ref{fig:f6} for discussion). We simulate a deep flow with a total depth of about 3.2 pressure scale height at latitude $\theta=90^{\circ}$, with the dimensionless rotation rate $\Omega$ spanning from 0 to 5.12, and the dimensionless total flux $F_{tot}$ spanning from $10^{-5}$ to $10^{-2}$. Additional parameters of the simulation cases are given in Table~\ref{table:tab1} (see Appendix \ref{appendixD}). The initial thermal structure of each simulation is a polytrope, with about 80 percent of the total flux $F_{tot}$ carried by radiation or conduction, and the rest 20 percent is carried by convection when convection is well developed (convective flux $F_{c}\sim 0.2 F_{tot}$). In Jupiter, almost all the flux is transported by convection. Here we raise the proportion of flux carried by radiation or conduction to keep the Prandtl number small
(the thermal conductivity $\kappa$ decreases more rapidly than the turbulent kinematic viscosity $\nu$ when flux is reduced, which leads to a raise of Prandtl number; note that scalings $\kappa \propto F_{tot}$ and $\nu\propto F_{tot}^{1/3}$ are approximately followed in large eddy simulations if grid resolution is fixed) and to speed up the thermal relaxation. This technique has been widely used in simulations of stellar and planetary convection \citep{brummell2002penetration,rogers2005penetrative,cai2020upward,cai2021deep}.In our numerical setting, the Ekman number has a scaling $E\propto\nu\propto F_{tot}^{1/3}$ and the flux-based Rayleigh number (see \citet{aurnou2020connections} for the definition) has a scaling $Ra_{F}\propto F_{tot}/(\nu \kappa^2)\propto F_{tot}^{-4/3}$. Therefore, $E$ decreases and $Ra_{F}$ increases with decreasing $F_{tot}$, respectively. Basically, the simulation reaches a parameter regime of higher $Ra_{F}$ and lower $E$ when $F_{tot}$ decreases. Previous simulations \citep{chan2007rotating,kapyla2011starspots,chan2013numerical,favier2014inverse,guervilly2014large,cai2021large} have revealed that the convective Rossby number $Ro_{c}$ plays an important role on the appearance of LSVs. In our cases, we compute $Ro_{c}=\left<v''\right>_{ref}/(2\Omega H)$, where $\left<v''\right>_{ref}$ is the averaged root-mean-square convective velocity calculated in an independent non-rotating reference case, $H$ is the height of the box, and $\Omega$ is the rotation rate. $\left<v''\right>_{ref}$ has been widely used as the characteristic velocity to evaluate $Ro_{c}$ in the astrophysical community \citep{kim1996theoretical,chan2001rotating,augustson2016magnetic}.  In the fluid dynamics community, the free-fall convective velocity is usually used as characteristic velocity to evaluate $Ro_{c}$ for Rayleigh-B\'enard convection. Simulations on Rayleigh-B\'enard convection \citep{cai2021large} have shown that the free-fall convective velocity is comparable to $\left<v''\right>_{ref}$ in magnitude. Therefore, there is no significant difference between using either of these two velocities for $Ro_{c}$.

\begin{figure*}[!htbp]
\includegraphics[width=\textwidth]{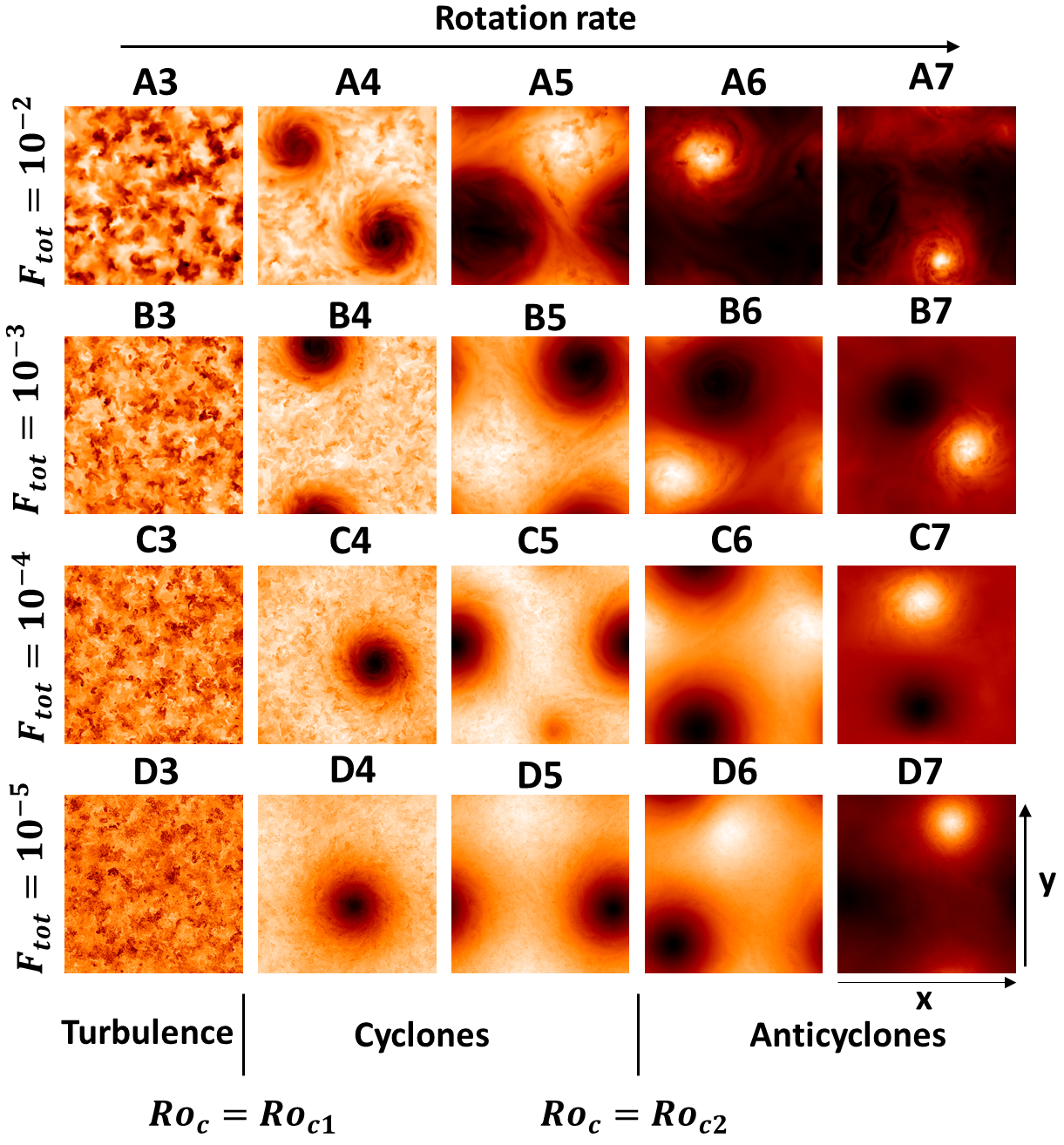}
\caption{Flow patterns for rotating convection at the high latitude $\theta=90^{\circ}$. Snapshots of the temperature perturbation taken horizontally at the mid of the convection zone $z=0.5$. The bright (dark) color denotes higher (lower) temperature. From top to bottom, the panels show four groups with different total fluxes $F_{tot}\in \{10^{-2},10^{-3},10^{-4},10^{-5}\}$, respectively. The subtitle of each small figure shows the specific case number. Flow patterns are separated into three regimes by $Ro_{c1}$ and $Ro_{c2}$: the turbulence, the large-scale cyclones, and the large-scale anticyclones. Note that the Rossby numbers (not the rotation rates) are close in each column. \label{fig:f1}}
\end{figure*}

After a small initial perturbation and a long period of evolution (at least several thermal relaxation time $\tau_{th}\sim 1/F_{tot}$), the convective flow reaches a statistically thermal equilibrium state. Two transitions in flow patterns take place as the rotation rate increases (Fig.~\ref{fig:f1}). The convective Rossby number $Ro_{c}$ , which measures the importance of buoyancy to rotational effects (see Appendix \ref{appendixB}), is an essential parameter to control the transitions of flow patterns \citep{chan2013numerical,guervilly2014large,cai2021large}.

\begin{figure*}[!htbp]
\includegraphics[width=\textwidth]{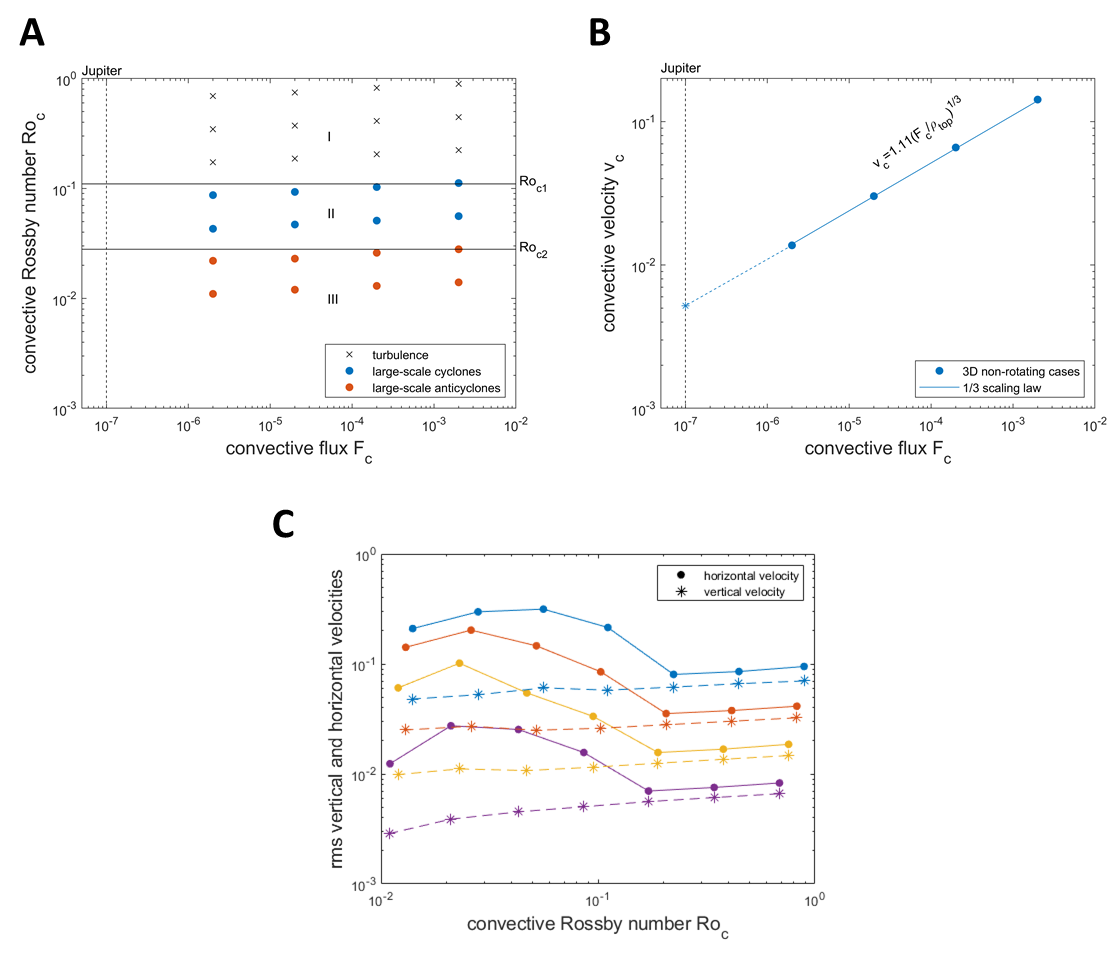}
\caption{Results of rotating flow at high latitudes. (A)Transitions of flow patterns in rapidly rotating convection. The computed cases are shown in the $F_{c}-Ro_{c}$ diagram. The symbols cross, blue cycle, and red cycle denote turbulence, large-scale cyclones, and the large-scale anticyclones, respectively. These three regimes I, II, and III are separated by $Ro_{c1}$ and $Ro_{c2}$. The vertical dashed line shows the location of Jupiter's convective flux in the diagram. (B)Scaling of convective velocity. The convective velocity $v_{c}$ as a function of the convective flux $F_{c}$ for the non-rotating Cases A0, B0, C0, and D0. The root-mean-square velocity $\left<v''\right>$ is used as $v_{c}$. $\rho_{top}$ is the density at the top of the box. (C) and (D) are the time averaged root-mean-square horizontal and vertical velocities, respectively. Groups A-D represent four groups with different total fluxes $F_{tot}\in \{10^{-2},10^{-3},10^{-4},10^{-5}\}$.
\label{fig:f2}}
\end{figure*}

The first transition from turbulence (regime I) to large-scale cyclones (regime II, the flow is still turbulent but large-scale cyclones appear) occurs when $ Ro_{c}$ is smaller than a critical number $Ro_{c1}\sim 0.11$ (Fig.~\ref{fig:f2}A). It is followed by a second transition to the appearance of large-scale anticyclones (regime III, the flow is still turbulent but large-scale anticyclones appear) when $Ro_{c}$ further decreases to another critical number $Ro_{c2}\sim 0.023$ (Fig.~\ref{fig:f2}A). This result is insensitive to various parameters, such as the magnitude of flux, the Prandtl number, and the Ekman number, at least in the parameter space that has been explored ($10^{-5}\leq F_{tot}\leq 10^{-3}$, $0.07\leq Pr \leq 8.1$, $2.5\times 10^{-6}\leq E\leq 1.8\times 10^{-4}$). The large-scale cyclones and anticyclones can survive for a long time. For example, the vortices in Case D7 are still alive after 1600 system rotation periods. In regime II, a pair of cyclones in Case A4 have survived for at least 100 system rotation periods till we terminated the simulation. It indicates that multiple cyclones could be generated in a deep rotating convection zone. When curvature effect is taken into account, these multiple cyclones tend to cluster around the pole \citep{cai2021deep}, which may explain the driving mechanics of the circumpolar cyclones observed in Jupiter's poles \citep{adriani2018clusters,adriani2020two}. In regime III, counter rotating vortex dipoles are formed in most cases, similar to the dipole configurations observed in the Jupiter's southern polar region ($10^{\circ}$ away from the pole) \citep{adriani2020two}. Mach number also has effect on the survivorship of cyclones. Case A6 shows that a cyclone (Fig.~\ref{fig:f1}) is twisted and almost destructed when the Mach number reaches one. Similar phenomenon is observed in Case A7. In other cases, cyclones are stable since their speeds are subsonic. Mach number decreases with the convective flux (Table~\ref{table:tab1}). The vertical velocity is much smaller than the horizontal velocity in rapidly rotating flow (Fig.\ref{fig:f2}C), thus the Mach number reflects the maximum wind speed of the LSVs. As the convective flux of Jupiter is lower, we expect that the LSVs in Jupiter are subsonic. The horizontal velocity reaches a peak value near the second transition state just before the large-scale anticyclones appear (Fig.~\ref{fig:f2}C). Passing through this transition state, the horizontal velocity and vortex size decrease with decreasing $Ro_{c}$. The sizes of LSVs are quantitatively measured and listed in the last column of Table~\ref{table:tab1}. The observed decreasing vortex size with decreasing $Ro_{c}$ is in agreement with the theoretical analysis of Coriolis-Inertial-Archimedean (\textbf{CIA}) balance on rapidly rotating convection \citep{christensen2010dynamo}, from which the relative size of LSVs at small $Ro_{c}$ obeys a scaling $\ell/h \sim Ro_{c}^{1/2}$, where $\ell$ and $h$ are horizontal and vertical length scales of a vortex respectively (see Appendix \ref{appendixB}).

Now we can apply the simulation result to Jupiter.  The convective flux $F_{c}$ calculated in the simulation group D ($F_{c}\sim 2\times 10^{-6}$) is close to the realistic values in gas giant planets (for Jupiter, $F_{c_J}\sim 1\times 10^{-7}$).
Our simulation has shown that the essential factor for determining the appearance of LSVs is the convective Rossby number $Ro_{c}$. It leads to the argument that this result could be extrapolated to the Jovian atmosphere. To obtain $Ro_{c}$ in Jupiter, we first make an estimation on the convective velocity $\left<v''\right>$ of a non-rotating Jupiter. From the mixing length theory, $\left<v''\right>$ has a scaling relation with the convective velocity $F_{c}$ of $\left<v''\right> \sim  ({F_{c}/\rho})^{1/3}$ \citep{bohm1958wasserstoffkonvektionszone,chan1989turbulent}, where $\rho$ is density. The $1/3$ scaling law is tested against the data of the non-rotating cases (Fig.\ref{fig:f2}B). The data shows a scaling relation $\left<v''\right>=1.11(F_{c}/\rho_{top})^{1/3}$, which is remarkably consistent with the theoretical mixing-length prediction. With this scaling relation, the dimensionless convective velocity of a non-rotating Jupiter is predicted to be 0.0052, which corresponds to $4.02 {\rm m/s}$. As the angular velocity of Jupiter is a constant $1.76\times10^{-4} {\rm rad/s}$, the appearances of large-scale cyclones or anticyclones in Jupiter solely depend on the depth of atmosphere $H$. For a given $Ro_{c1}$ or $Ro_{c2}$, the lower limit of $H$ can be inferred. Recently, large-scale cyclones and anticyclones have been observed on Jupiter's poles by the Juno spacecraft \citep{adriani2018clusters,adriani2020two}. The condition for the emergence of LSVs can be applied to the poles. To satisfy $Ro_{c}\leq Ro_{c1}$, we expect that the depth of Jovian atmosphere is deeper than $H_{c1}=\left<v''\right>_{ref}/(2\Omega Ro_{c1})\approx 102 {\rm km}$ in the polar region. To generate anticyclone in the polar region, it is expected that the Jovian atmosphere is deeper than $H_{c2}=\left<v''\right>_{ref}/(2\Omega Ro_{c2})\approx 501 {\rm km}$. As both cyclones and anticyclones exist, we predict that Jupiter's atmosphere is at least $500 {\rm km}$ deep in the polar regions. Juno's gravitational measurement predicted that the depth of atmosphere at polar regions is in the range of $0\sim 500 {\rm km}$ \citep{kaspi2018jupiter}. The lower limit of our prediction coincides with the upper limit of the prediction from gravitational measurement.

\subsection{How to generate GRS-like LSVs in low latitudes?}
We have shown that LSVs can be generated in a rapidly rotating deep convective flow at high latitudes. Now we demonstrate by simulations that LSVs can also survive in a rotating flow at low latitudes. For simulation parameters at low latitudes, $\theta$ is fixed at $22.5^{\circ}$, $F_{tot}$ is from $10^{-5}$ to $10^{-3}$, and $\Omega$ is from 0.16 to 2.56 (see Table~\ref{table:tab2} in Appendix \ref{appendixD}). The basic settings are almost the same as the simulations at high latitudes, except that now the rotational axis is inclined with gravitational axis by an angle $67.5^{\circ}$.

\begin{figure*}[!htbp]
\centering
\includegraphics[width=0.85\textwidth]{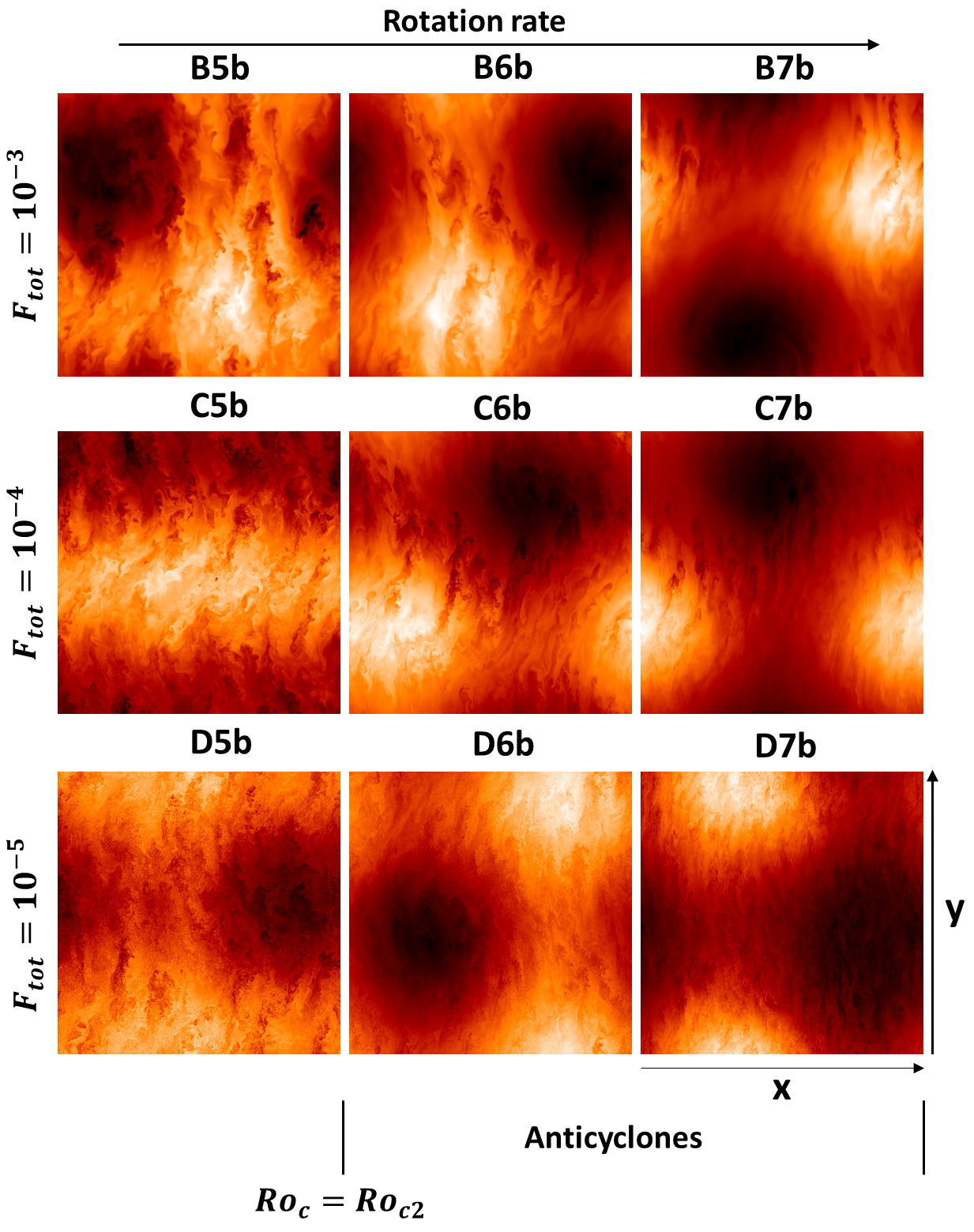}
\caption{Flow patterns for rotating convection at the low latitude $\theta=22.5^{\circ}$. Snapshots of the temperature perturbation taken horizontally at the mid of the convection zone $z=0.5$. The rotation rate increases from left to right. Three groups with different total fluxes $F_{tot} \in \{10^{-2},10^{-3},10^{-4}\}$ are shown from top to bottom. The settings are almost identical to B5-D7 in Fig.~\ref{fig:f1}, except that the latitude is different.\label{fig:f3}}
\end{figure*}

\begin{figure*}[!htbp]
\centering
\includegraphics[width=0.85\textwidth]{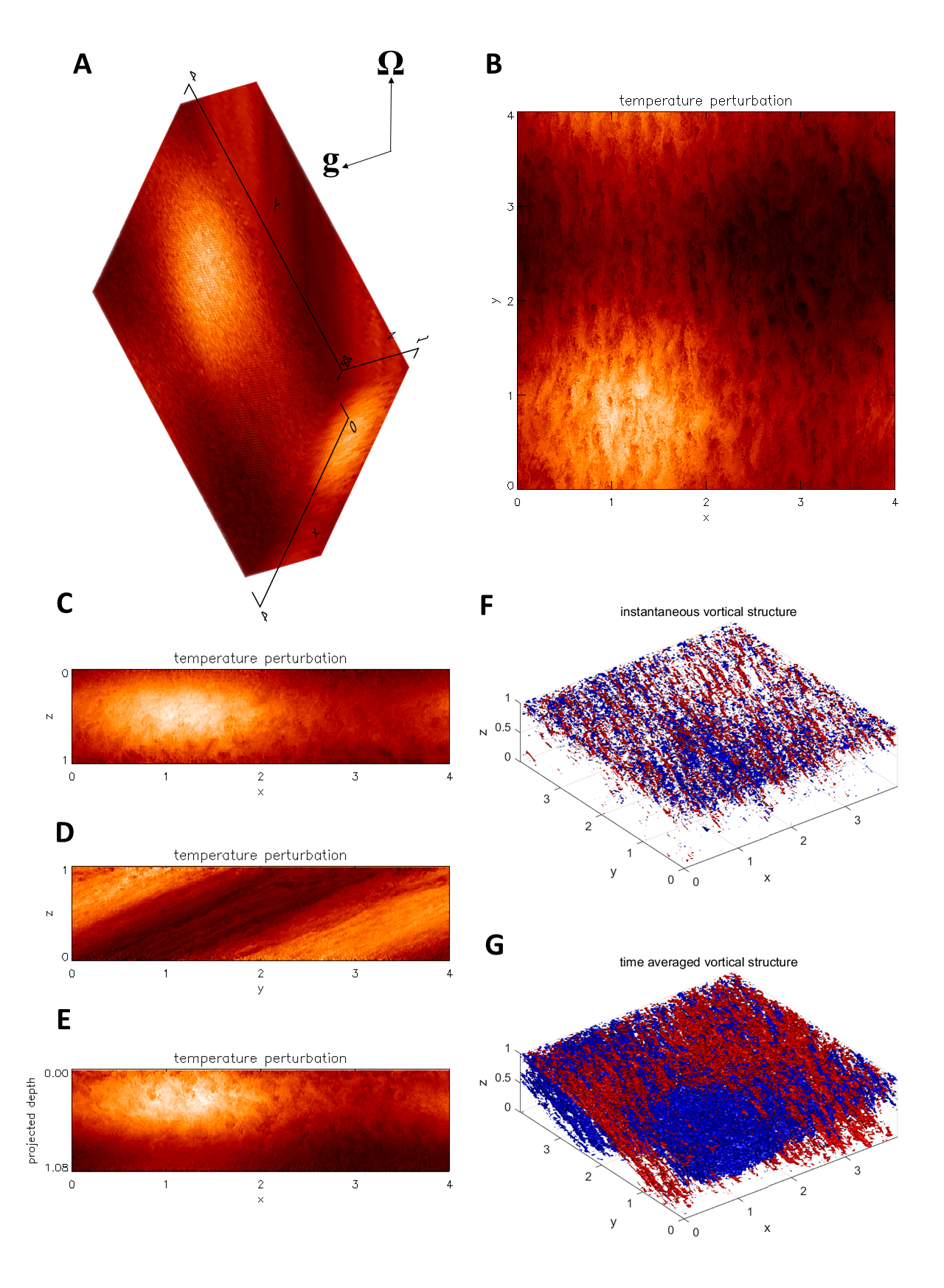}
\caption{(A-E)Snapshots of temperature perturbation of Case D7b. (A)The 3D volume render. (B)Contours at the plane $z=0.9$. (C)Contours at the plane $y=3.6$. (D)Contours at the plane $x=1.0$. (E)Contours at a plane perpendicular to the rotational axis. (F)The vortical structure of an instantaneous velocity field. (G)The vortical structure of time averaged (about 27 system rotation periods) velocity field. Red (blue) represents positive (negative) vertical component of vorticity. The vortical structure is identified by $\lambda_{2}$ criterion (see Appendix \ref{appendixC}).   \label{fig:f4}}
\end{figure*}

\begin{figure*}[!htbp]
\centering
\includegraphics[width=\textwidth]{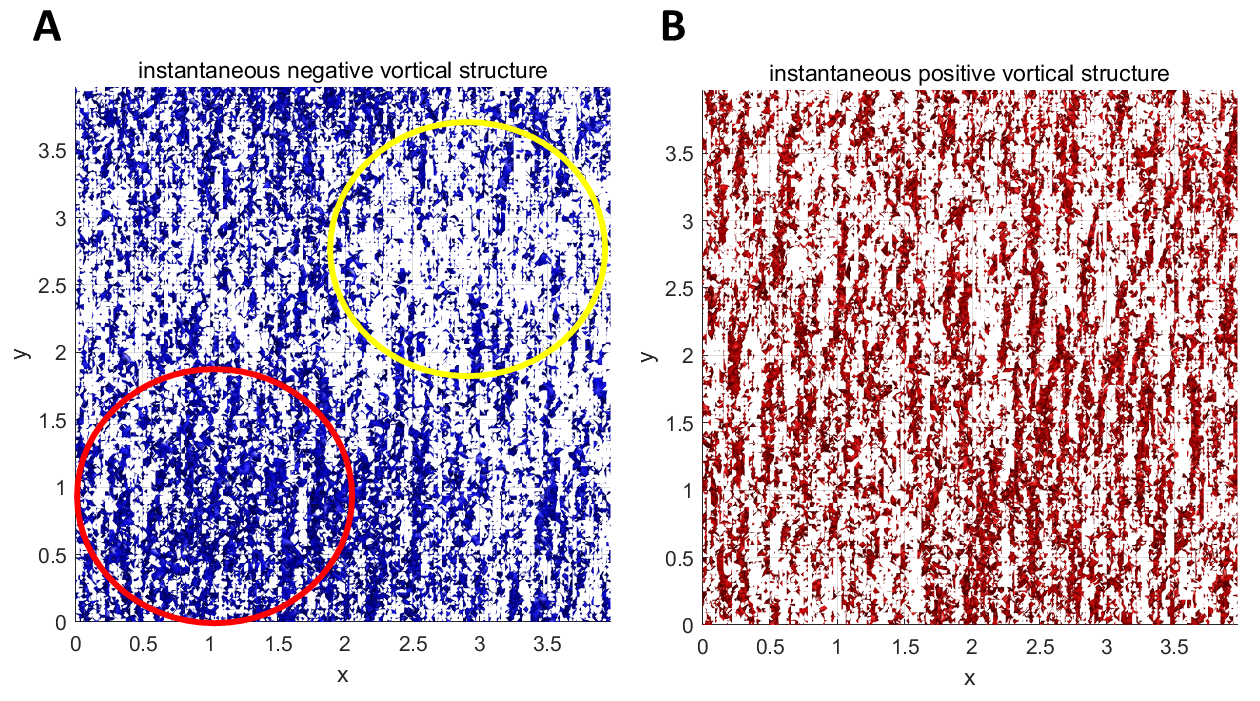}
\caption{Vortical structures of Case D7b viewed from the above. (A)The vortical structure with negative vertical vorticity. The red circle shows the location of large-scale anticyclone, and the yellow circle shows the location of large-scale cyclone. (B)The vortical structure with positive vertical vorticity. \label{fig:s1}}
\end{figure*}

The appearances of LSVs in rapidly rotating flow at low latitudes are shown in Fig.~\ref{fig:f3}. Also apparent is that large scale anticyclone, similar to GRS observed at Jupiter's low latitude, are formed when $Ro_{c}$ is small. In simulations at high latitudes, columnar circular vortices are aligned with the rotational axis. Unlike previous simulations, here we see that the axially aligned columnar vortices are elliptic on the plane perpendicular to $\boldsymbol{\Omega}$ rather than circular (Fig.~\ref{fig:f4}A). This can be shown more clearly by showing the flow structure at a cut plane perpendicular to the rotational axis (Fig.~\ref{fig:f4}E). It demonstrates that elliptical vortex could be generated in deep convective rotating flow. The flow patterns on the meridional-zonal ($x$-$y$), zonal-vertical ($x$-$z$), and meridional-vertical ($y$-$z$) planes can be viewed as the projections of the elliptical vortex. Interestingly, the large-scale anticyclone on the $x$-$y$ plane shows a circular structure (Fig.~\ref{fig:f4}B), while on the $x$-$z$ plane it remains elliptic (Fig.~\ref{fig:f4}C). Recently, observation has shown that the GRS is changing from elliptical shape to circular shape \citep{simon2018historical}, which shares some similarity with our simulation result. Tilted patterns, which are parallel to $\bm{\Omega}$ as exhibited on the $y$-$z$ plane (Fig.4D), demonstrating that the columnar structure is indeed aligned with $\bm{\Omega}$. As indicated in the Taylor-Proudman theory, it is not surprising to observe that the columnar structure is aligned with the rotation axis. In a rapidly rotating system, the vortices are expected to form parallel to the rotation axis regardless of the simulation domain geometry and latitude.
The LSVs are stable for a very long time. For example, the vortices in Case D7b have survived for at least 400 rotating periods till we terminate the simulation. Movies S1, S2, S3 and S4 (see Appendix \ref{appendixE}) show the flow motions at different planes $z=0.1$, $z=0.9$, $y=3.6$, and $x=1.0$ for about 27 system rotating periods. In movies S1 (horizontal plane cut at z=0.1) and S2 (at z=0.9), small scale turbulent motions are observed inside the anticyclone, similar to the fluid motions observed in Juno images \citep{sanchez2018rich}. The large-scale anticyclone can be better understood by showing vortical structures. Surprisingly, small vortical tubes (SVTs) (aligned with $\bm{\Omega}$) are observed in the instantaneous vortical structure (Fig.\ref{fig:f4}F). Tube-like vortices were found associated with strong down flows in deep convection \citep{brummell2002penetration}. Movie S5 on vortical structure shows that LSVs are associated with the organized movements of these SVTs. The distributions of positive and negative SVTs are asymmetric. The positive SVTs are almost evenly distributed in the fluid domain (Fig.~\ref{fig:s1}B). However, the distribution of negative SVTs are uneven (Fig.~\ref{fig:s1}A), with more distributed inside the anticyclone and less distributed inside the cyclone. This is further demonstrated in Fig.~\ref{fig:f4}G, which clearly shows the large-scale vortical columns in the time averaged vortical structure.

There are a number of processes that are responsible for causing the asymmetric distributions of positive/negative SVTs in the positive/negative LSVs. First, both laboratory experiments \citep{vorobieff2002turbulent} and numerical simulations \citep{guervilly2014large,cai2021large} of rotating convection have shown that ejection of thermal plumes is more frequent than injection at thermal boundary layers. The more frequent ejection of thermal plumes, the more positive SVTs are created. Second, the rotational effects are different in the cyclones and anticyclones. Cyclone spins in the same direction as the system rotation, while anticyclone spins in the opposite way. Thus, the effective rotation rate (the summation of the spinning rate of the LSV and the system rotation rate) is larger inside a cyclone than an anticyclone. As a result, the suppressing effect of rotation on convection is stronger inside a cyclone than an anticyclone \citep{chan2013numerical}. The suppressing effect reduces the numbers of both injected and ejected plumes, but more are reduced inside the cyclone. As the number of injected plumes (negative SVTs) is fewer than that of ejected plumes (positive SVTs), a significant reduction of injected plumes can lead to the sparse distribution of negative SVTs inside the cyclone. Third, two like-signed vortices tend to rotate around each other when they are strong and close enough \citep{boubnov1986experimental,hopfinger1993vortices}. The patch of the two like-signed vortices has stronger intensity, and can attract more like-signed vortices to rotate together \citep{yasuda1995two}. Finally, LSVs can be formed by the clustering of these like-signed vortices. Inside the large-scale cyclone, the intensities of positive SVTs are stronger and their distributions are denser than those of negative SVTs. The clustering of these strong positive SVTs expels negative SVTs to form an anticyclonic region \citep{guervilly2014large}. When the negative SVTs are strong enough and their distances are short enough, a large-scale anticyclone can be formed in the anticyclonic region.

Compared with the cases at high latitudes, the sizes of LSVs at low latitudes are larger. Since the dominant size of LSVs is in the $x$-$y$ plane, we expect that the vortex size follows the relation $\ell/h \sim [\left<v''\right>_{ref}/(2\Omega H \sin\theta )]^{1/2}=(Ro_{c}/\sin\theta)^{1/2}$ at the low Rossby regime, and thus, the vortex size at $\theta=22.5^{\circ}$ is about 1.6 times larger than that at $\theta=90^{\circ}$.
To fit LSVs in the box at a small $\theta$, it requires that either the box size is wide enough or $Ro_{c}$ is small enough. Otherwise, only vague vortices or shear flows exist (the first column of Fig.\ref{fig:f3}). The shear flows might be closely related to the strong prograde and retrograde jets observed in Jupiter's equatorial region. Near the equatorial region, $\theta$ is very small. Since the vortex size is proportional to $(1/\sin\theta)^{1/2}$, LSVs are unlikely to be formed in the meridional-zonal plane. Alternatively, shear flows will be favored in this region. We expect that, if the box is large enough, the critical Rossby numbers for the appearances of LSVs would remain the same as at high latitudes. Thus we predict that the Jovian atmosphere at low latitudes is deeper than 500km.

\section{Discussion}
Our results suggest that long-lived LSVs could be generated in a rotating deep convective flow at both high and low latitudes, when the convective Rossby number $Ro_{c}$ is smaller than certain critical values. In particular, anticyclonic LSVs similar to the GRS can be found in low latitudes for certain critical values of this number. We find that there are positive and negative SVTs inside the LSVs. The distributions of negative SVTs are nonuniform inside the LSVs, with more distributed inside the large-scale anticyclone and less distributed inside the large-scale cyclone. We have proposed three mechanisms to explain it. First, the preference of ejection of thermal plumes in the thermal boundary layers tend to create more positive SVTs than negative ones. Second, the effective rotation rate is larger in the large-scale cyclone than anticyclone, which reduces more negative SVTs in the large-scale cyclone. Third, the clustering of positive SVTs expels negative SVTs in the cyclone, but the clustering of negative SVTs is not strong enough to significantly expel positive SVTs in the anticyclone. We also found that the critical values of $Ro_{c}$ for the appearances of LSVs are insensitive to other parameters, such as the magnitude of normalized flux and the Ekman number. This finding may have certain implication in understanding astrophysical or geophysical flows. In stars or planets, the normalized flux and Ekman number are usually too small to simulate directly. The convective Rossby number, a viscous-free parameter, however, is not as small as the normalized flux and Ekman number, and may be reachable in numerical simulations. It can be anticipated that using similar values of convective Rossby number in the simulations of stars or planets may provide some insights on understanding the dynamics of these astrophysical or geophysical flows. For the Sun, one can estimate that $Ro_{c}$ is in the range of 0.02 to 0.2. The estimation is based on the physical values that the depth of the solar convection zone $H\approx 2.1\times 10^{8} m$, the rotation rate $\Omega\approx 2.6\times 10^{-6} s^{-1}$, and the convective velocity $\left<v''\right>\approx 20-200m/s$ \citep{miesch2008structure}. On the basis of our calculation, we speculate that LSVs might exist at the bottom of solar convection zone, where the convective Rossby number is at around 0.02. For fast rotating solar-like stars, such as HII 296, BD-16351, and TYC 5164-567-1 \citep{folsom2016evolution}, the convective Rossby numbers are about 10 times smaller than that of Sun. Therefore, LSVs could probably be formed in the surface convection zones of these solar type stars. However, these estimations do not consider the effect of magnetic field. It has been found that a strong magnetic field can inhibit the formation of LSVs in rapidly rotating convection \citep{guervilly2015generation}. Application of the result to stars requires further investigation on the effect of magnetic field.

%Simulations at low latitudes indicate that the columnar LSVs tend to be aligned with the rotational axis (analogy to Taylor-Proudman theorem), rather than the radial direction. If the GRS is formed by rotating turbulent thermal convection, we conjecture that it may also have such a columnar structure, which can extend at least 500 km deep into the atmosphere. The microwave radiometer in the Juno spacecraft can see up to 350 km deep into the GRS in Jupiter. The conjecture can be tested against the microwave radiometer data in the future. The Juno gravity measurements can also provide information on the depth of GRS \citep{parisi2020mascon,galanti2019determining} when more data are available in the future.

The microwave radiometer in the Juno spacecraft can see up to 350 km deep into the GRS in Jupiter. The Juno gravity measurements can also provide information on the depth of GRS \citep{parisi2020mascon,galanti2019determining}. Recently, the Juno microwave radiometer data \citep{bolton2021microwave} and the gravity data \citep{parisi2021depth} reveal that the GRS has a root which could possibly extend down to 500 $km$ deep. It has to be mentioned that our model is based on the theory of rotating convection on a $f$-plane. We did not exclude the possibility that LSVs could be formed by other mechanisms, such as the formation of LSVs by opposite zonal jets \citep{lemasquerier2020remote} where LSVs can be formed in a shallower atmosphere. Also, we did not include the curvature $\beta$-effect in our model. When $\beta$-effect is taken into account, the conditions for the appearance of LSVs at low latitudes could be more restricted. It is challenging to set an appropriate boundary condition in the meridional direction for a box simulation if $\beta$-effect and zonal jets are taken into account. Global simulations \citep{cai2012three,yadav2020deepb} will be more suitable for such kind of problems. We defer the discussion of global simulations of LSVs to future works.

\begin{acknowledgements}
We thank the reviewer for the helpful comments on the manuscript. This work was partially supported by NSFC (Nos.~12173105,11503097), the Science and Technology Development Fund, Macau SAR (No.~0156/2019/A3), the Guangdong Basic and Applied Basic Research Foundation (No.~2019A1515011625), the China Space Agency Project (No.~D020303), and the China Manned Space Project (No.~CMS-CSST-2021-B09). Part of the simulation was performed on the supercomputers at the National Supercomputer Center in Guangzhou.
\end{acknowledgements}

\appendix

\section{Numerical Method}\label{appendixA}
For a rotating compressible flow in a Cartesian box, the hydrodynamic equations of mass, momentum, and energy conservations are
\begin{eqnarray}
\partial_{t}\rho&=&-\mathbf{\nabla}\cdot \mathbf{M} ~,\\
\partial_{t}\mathbf{M}&=&-\mathbf{\nabla}\cdot (\mathbf{MM}/\rho)+\mathbf{\nabla}\cdot \mathbf{\Sigma}-\mathbf{\nabla} p+\rho \mathbf{g}-2\mathbf{\Omega}\times \mathbf{M}~,\\
\partial_{t}E&=&-\nabla \cdot[(E+p)\mathbf{M}/\rho-\mathbf{M}\cdot\mathbf{\Sigma}/\rho+\mathbf{F_{d}}]+\mathbf{M}\cdot \mathbf{g} ~.
\end{eqnarray}
where $\rho$ is density, $\mathbf{M}$ is momentum, $\mathbf{\Sigma}$ is stress tensor, $p$ is pressure, $\mathbf{g}$ is gravity, $\mathbf{\Omega}$ is angular velocity, $E$ is total energy, $\mathbf{F_{d}}$ is diffusive or conductive flux. We solve the hydrodynamic equations by a mixed finite-difference spectral model \citep{cai2016semi}. This method has been used in the study of compressible turbulent convection \citep{cai2018numerical} and convective overshooting \citep{cai2020upward}. In this paper, additional Coriolis term has been added in the momentum equation. In the model, all the linear terms, including the Coriolis term, are integrated semi-implicitly. An advantage of this model is that the numerical time step is not restricted by the CFL conditions associated with speeds of acoustic wave, gravity wave, or inertial wave. Thus the model is suitable for simulating rapidly rotating compressible turbulent convection at small flux. In this paper, the initial thermal structure is assumed to be in a polytropic state:
\begin{eqnarray}
T/T_{top} &=& 1+\eta (1-z)~, \\
\rho/\rho_{top} &=& (T/T_{top})^{m}~, \\
p/p_{top} &=& (T/T_{top})^{m+1}~,
\end{eqnarray}
where $T$ is the temperature, $\eta$ is a ratio measuring the depth of fluid, $m$ is the polytropic index, and the subscript $top$ denotes the corresponding value at the top of the computational domain. All the physical variables are normalized by the height, density, pressure, and temperature at the top. The adiabatic polytropic index is set to be $m_{ad}=1.5$. The fluid is convectively unstable when $m<m_{ad}$, and stable when $m>m_{ad}$. In our settings, we consider the rapidly rotating convection in a pure convection zone with $\eta=4$ and $m=1$. The numerical experiments are performed with a large eddy simulation method, in which the small scale effects are modelled by a sub-grid-scale (SGS) turbulent model. In the SGS model \citep{smagorinsky1963general}, the turbulent kinematic viscosity is evaluated by
\begin{equation}
\nu=c_{d}^2 (\Delta x\Delta y)^{1/2}\Delta z (2\mathbf{\sigma:\sigma})^{1/2}~,
\end{equation}
where $c_{d}=0.28$ is the Deardorff coefficient, $\mathbf{\sigma}$ is the strain rate tensor, and $\Delta x$, $\Delta y$, and $\Delta z$ are the mesh grid sizes in the $x$-, $y$-, and $z$-directions, respectively. The turbulent viscosity is decomposed into two components: a time-dependent horizontal mean component and a fluctuating component. In the numerical calculations, the higher order viscous terms associated with the fluctuating component are disregarded. Impenetrable and stress-free boundary conditions are used for velocities at the top and bottom of the box. The temperature is set to be a constant at the top, and a constant flux $F_{tot}=\kappa \eta$ is supplied at the bottom. Here the thermal conductivity $\kappa$ has a constant value throughout the domain. In the horizontal directions, periodic boundary conditions are used for all the thermodynamic variables. The computational domain is a {\it f}-plane with vertical {\it z}-direction inclined to the rotational axis by an angle $90^{\circ}-\theta$, where $\theta$ is the latitude. The horizontal {\it x}- and {\it y}-directions of the {\it f}-plane are corresponding to the west-to-east zonal and south-to-north meridional directions in the spherical coordinates, respectively. Fourier spectral expansions are used in the horizontal directions, and finite difference method is used in the vertical direction. We use density $\rho$, pressure $p$, total energy $E$, divergence of horizontal momentum $\delta=\bm{\nabla}_{H} \bm{\cdot} \bm{M}_{H}$, vertical component of the curl of horizontal momentum $\xi =\hat{\bm{z}}\bm{\cdot} \bm{\nabla}_{H} \bm{\times} \bm{M}_{H}$ as prognostic variables. The components of the horizonal momentums $\bm{M}_{H}=(M_{x},M_{y})$ are evaluated by linear combinations of $\delta$ and $\xi$. In the {\it f}-plane, additional terms on the horizonal means of $M_{x}$ and $M_{y}$ are required, which can be integrated from the following equations
\begin{eqnarray}
&&\partial_{t}\overline{M_{x}}=-\partial_{z}(\overline{M_{x}M_{z}/\rho})+\partial_{z}(\mu\partial_{z} \overline{M_{x}/\rho})+2\Omega_{z} \overline{M_{y}}~,\\
&&\partial_{t}\overline{M_{y}}=-\partial_{z}(\overline{M_{y}M_{z}/\rho})+\partial_{z}(\mu\partial_{z} \overline{M_{y}/\rho})-2\Omega_{z} \overline{M_{x}}~.
\end{eqnarray}
where $\mu$ is the dynamic viscosity, $\Omega_{z}$ is the vertical component of $\bm{\Omega}$, and the overline denotes horizontal average.

\section{Scalings of CIA Balance}\label{appendixB}
For a compressible flow, the vorticity equation can be obtained by taking curl on both sides of the moment equation:
\begin{eqnarray}
\partial_{t}\bm{\omega}+&(\bm{u}\bm{\cdot}\bm{\nabla}) \bm{\omega}-(\bm{\omega}\bm{\cdot}\bm{\nabla}) \bm{u}
=2\bm{\Omega}\bm{\cdot} \bm{\nabla} \bm{u}-2\bm{\Omega} \bm{\nabla}\bm{\cdot} \bm{u}+\bm{\nabla} \bm{\times} (\frac{1}{\rho}\bm{\nabla}\bm{\cdot}\bm{\Sigma})-\bm{\nabla} \bm{\times} (\frac{1}{\rho}\bm{\nabla}p')+\bm{\nabla} \bm{\times} (\frac{\rho'}{\rho} \bm{g})~,
\end{eqnarray}
where $\bm{u}$ is the velocity, $\bm{\omega}=\bm{\nabla}\bm{\times}\bm{u}$ is the vorticity, $\bm{\Sigma}$ is the stress tensor, and $p'$ and $\rho'$ are perturbations from their equilibrium states. In rapidly rotating convection, \textbf{CIA} balance among the Coriolis ({\textbf{C}}) term, the inertial (\textbf{I}) term, and the Archimedean (\textbf{A}) buoyancy term is achieved in the vorticity equation \citep{christensen2010dynamo}:
\begin{eqnarray}
2\bm{\Omega}\bm{\cdot} \bm{\nabla} \bm{u} \sim (\bm{u}\bm{\cdot}\bm{\nabla}) \bm{\omega} \sim \bm{\nabla} \bm{\times} (\frac{\rho'}{\rho} \bm{g})~.
\end{eqnarray}
A scale analysis of the terms in the \textbf{CIA} balance gives \citep{aurnou2020connections}
\begin{eqnarray}
\frac{2\Omega U}{h} \sim \frac{U^2}{\ell^2} \sim \frac{U^2_{c}}{\ell^2}~,
\end{eqnarray}
where $U$ is the characteristic velocity of the LSVs, $U_{c}$ is the characteristic velocity of non-rotating convection, and $h$ and $\ell$ are the vertical and horizontal length scales of LSVs, respectively. It quickly obtains
\begin{eqnarray}
Ro_{c}=\frac{U_{c}}{2\Omega h} \sim (\frac{\ell}{h})^2~.
\end{eqnarray}
As a result, the relative size of LSVs at small $Ro_{c}$ obeys a scaling $\ell/h \sim Ro_{c}^{1/2}$. It is consistent with the prediction of \citet{aurnou2020connections}, where they used the terminology system-scale Rossby number.

\section{Extraction of Vortices}\label{appendixC}
We extract vortices by the $\lambda_{2}$-criterion \citep{jeong1995identification} that the quantity $\lambda_{2}$ is defined as the second largest eigenvalue of the $3\times 3$ martrix $\bm{A}=\bm{S}^2+\bm{\omega}^2$, where $\bm{S}=(\bm{J}+\bm{J}^{T})/2$ and $\bm{\omega}=(\bm{J}-\bm{J}^{T})/2$, and the velocity gradient tensor $\bm{J}=\bm{\nabla}\bm{u}$. The vortex shape is then identified by choosing a specific negative value of $\lambda_{2}$ and plotting the corresponding isosurface.

\section{Simulation Parameters}\label{appendixD}
To explore how flux affects the flow pattern in rapidly rotating flow, we performed simulations on four groups of fluxes  with dimensionless values $F_{tot}\in \{10^{-2}, 10^{-3}, 10^{-4}, 10^{-5}\}$ at latitude $\theta=90^{\circ}$. We then choose eight different angular velocities $\Omega$ in each group to investigate the effect of rotation, giving a total of 32 simulation cases. The non-rotating cases are computed for reference. The detailed parameters of each simulation case are given in Table~\ref{table:tab1}. $F_{tot}$ is the total flux (about 80\% of $F_{tot}$ is transported by adiabatic temperature gradient); $\Omega$ is the angular velocity; $v''$ is the root-mean-square (rms) velocity; $v_{z}''$ is the rms vertical velocity; $v_{h}''$ is the rms horizontal velocity; ${Pr}=\langle \rho c_{p}\nu/\kappa\rangle$ is the averaged Prandtl number, where $c_{p}$ is the specific heat capacity at constant pressure, $\nu$ is the kinematic viscosity, and $\kappa$ is the thermal conductivity; ${Re}=\langle v'' H/\nu\rangle$ is the averaged Reynolds number; ${Re_{z}}=\langle v_{z}'' H/\nu\rangle$ is the averaged vertical Reynolds number; ${Ma}=\overline{\max(v/c_{s})}$ is the time averaged maximum Mach number at the top layer; ${Ro_{c}}=\langle v''/(2\Omega H)\rangle$ is the averaged convective Rossby number; ${E}=\langle \nu/2 \Omega H^2\rangle$ is the averaged Ekman number. The symbol $\langle \rangle$ denotes that the average is taken both temporarily and spatially throughout the computational domain. The symbol overline denotes that the average is taken temporarily. The symbols `{\it n}', `{\it c}', and `{\it ac}' denote no LSV, large-scale cyclone, and large-scale anticylone, respectively. The last column gives the corresponding radii of LSVs. The radius of a LSV is measured by the distance from the center of the LSV to the maximum value of averaged tangential velocity around the center at the plane $z=0.5$.
\startlongtable
\begin{deluxetable*}{ccccccccccccccc}
\tablecaption{Parameters of simulation cases of {\it f}-plane at high latitude $\theta=90^{\circ}$\label{table:tab1}}
\tablehead{
 Case & $F_{tot}$ & $\Omega$ & $\langle v''\rangle$ & $\langle v_{z}''\rangle$ & $\langle v_{h}''\rangle$ & $ Pr$ & $ Re$ & $ Re_{z}$ & $Ma$ & $Ro_{c}$ & $ E$ & {LSV} & Radius
}
\startdata
A0 &   $1.0\times 10^{-2}$  & 0    & 0.143 & 0.075 & 0.119 & 0.084 & 5036 & 2614  & 0.413 & $\infty$ & $\infty$ & {\it n} & -\\
A1 &   $1.0\times 10^{-2}$  & 0.08 & 0.120 & 0.070 & 0.095 & 0.082 & 4179 & 2425  & 0.384 & 0.891 & $1.8\times 10^{-4}$ & {\it n} & -\\
A2 &   $1.0\times 10^{-2}$  & 0.16 & 0.109 & 0.066 & 0.085 & 0.080 & 3775 & 2264  & 0.377 & 0.446 & $9.0\times 10^{-5}$ & {\it n} & -\\
A3 &   $1.0\times 10^{-2}$  & 0.32 & 0.103 & 0.062 & 0.081 & 0.079 & 3491 & 2102  & 0.382 & 0.223 & $4.6\times 10^{-5}$  & {\it n} & -\\
A4 &   $1.0\times 10^{-2}$  & 0.64 & 0.223 & 0.058 & 0.214 & 0.077 & 8301 & 2002  & 0.465 & 0.111 & $2.3\times 10^{-5}$  & {\it c} & 0.36 \\
A5 &   $1.0\times 10^{-2}$  & 1.28 & 0.321 & 0.061 & 0.315 & 0.075 & 12229 & 2141 & 0.699 & 0.056 & $1.1\times 10^{-5}$ & {\it c}  & 1.00\\
A6 &   $1.0\times 10^{-2}$  & 2.56 & 0.303 & 0.052 & 0.297 & 0.066 & 13224 & 2125 & 1.304 & 0.028 & $4.9\times 10^{-6}$ & {\it ac} & 0.58 \\
A7 &   $1.0\times 10^{-2}$  & 5.12 & 0.216 & 0.048 & 0.210 & 0.069 & 9102 & 1907  & 1.154 & 0.014 & $2.5\times 10^{-6}$ & {\it ac} & 0.34\\
\hline
B0 &   $1.0\times 10^{-3}$ &  0    & 0.066 & 0.034 & 0.055 & 0.396 & 4703 & 2472  & 0.195 & $\infty$ & $\infty$ & {\it n} & -\\
B1 &   $1.0\times 10^{-3}$ &  0.04 & 0.053 & 0.032 & 0.041 & 0.395 & 3772 & 2299  & 0.186 & 0.825    & $1.8\times 10^{-4}$ & {\it n} & -\\
B2 &   $1.0\times 10^{-3}$ &  0.08 & 0.049 & 0.030 & 0.037 & 0.379 & 3436 & 2126  & 0.186 & 0.413    & $8.9\times 10^{-5}$ & {\it n} & -\\
B3 &   $1.0\times 10^{-3}$ &  0.16 & 0.045 & 0.028 & 0.035 & 0.373 & 3182 & 1969  & 0.188 & 0.206    & $4.5\times 10^{-5}$ & {\it n} & -\\
B4 &   $1.0\times 10^{-3}$ &  0.32 & 0.089 & 0.026 & 0.085 & 0.367 & 6822 & 1848  & 0.217 & 0.103    & $2.2\times 10^{-5}$ & {\it c} & 0.38\\
B5 &   $1.0\times 10^{-3}$ &  0.64 & 0.149 & 0.025 & 0.146 & 0.354 & 11999 & 1845  & 0.274 & 0.052   & $1.1\times 10^{-5}$ & {\it c} & 0.62\\
B6 &   $1.0\times 10^{-3}$ &  1.28 & 0.205 & 0.027 & 0.203 & 0.342 & 17196 & 2090  & 0.547 & 0.026   & $5.3\times 10^{-6}$ & {\it c}+{\it ac} & 0.52+0.50\\
B7 &   $1.0\times 10^{-3}$ &  2.56 & 0.144 & 0.025 & 0.142 & 0.341 & 12069 & 1918  & 0.707 & 0.013   & $2.6\times 10^{-6}$ & {\it c}+{\it ac} & 0.46+0.38\\
\hline
C0 &   $1.0\times 10^{-4}$ &  0    & 0.030 & 0.016 & 0.025 & 1.844 & 4556 & 2430  & 0.091 &$\infty$ & $\infty$ & {\it n} & -\\
C1 &   $1.0\times 10^{-4}$ &  0.02 & 0.024 & 0.015 & 0.019 & 1.792 & 3594 & 2215  & 0.087 & 0.756    & $1.7\times 10^{-4}$ & {\it n} & -\\
C2 &   $1.0\times 10^{-4}$ &  0.04 & 0.022 & 0.014 & 0.017 & 1.760 & 3259 & 2046  & 0.087 & 0.378    & $8.4\times 10^{-5}$ & {\it n} & -\\
C3 &   $1.0\times 10^{-4}$ &  0.08 & 0.020 & 0.013 & 0.016 & 1.739 & 3004 & 1879  & 0.087 & 0.189    & $4.2\times 10^{-5}$ & {\it n} & -\\
C4 &   $1.0\times 10^{-4}$ &  0.16 & 0.035 & 0.012 & 0.033 & 1.726 & 5681 & 1730  & 0.101 & 0.095    & $2.1\times 10^{-5}$ & {\it c} & 0.34\\
C5 &   $1.0\times 10^{-4}$ &  0.32 & 0.056 & 0.011 & 0.055 & 1.711 & 9190 & 1626  & 0.114 & 0.047    & $1.1\times 10^{-5}$ & {\it c} & 0.44\\
C6 &   $1.0\times 10^{-4}$ &  0.64 & 0.102 & 0.011 & 0.102 & 1.780 & 16330 & 1620  & 0.183 & 0.023  & $5.5\times 10^{-6}$ & {\it c}+{\it ac} & 0.66+0.52\\
C7 &   $1.0\times 10^{-4}$ &  1.28 & 0.062 & 0.099 & 0.061 & 1.743 & 9954 & 1464  & 0.246 & 0.012  & $2.7\times 10^{-6}$ & {\it c}+{\it ac} & 0.32+0.46\\
\hline
D0 &   $1.0\times 10^{-5}$ &  0    & 0.014 & 0.007 & 0.011 & 8.617 & 4399 & 2350  & 0.043 & $\infty$ & $\infty$ & {\it n} & -\\
D1 &   $1.0\times 10^{-5}$ &  0.01 & 0.011 & 0.007 & 0.008 & 8.371 & 3421 & 2127  & 0.040 & 0.686 & $1.6\times 10^{-4}$ & {\it n} & -\\
D2 &   $1.0\times 10^{-5}$ &  0.02 & 0.098 & 0.006 & 0.008 & 8.252 & 3102 & 1952  & 0.040 & 0.343 & $8.0\times 10^{-5}$  & {\it n} & - \\
D3 &   $1.0\times 10^{-5}$ &  0.04 & 0.009 & 0.006 & 0.007 & 8.170 & 2845 & 1786  & 0.040 & 0.171 & $4.0\times 10^{-5}$ & {\it n} & - \\
D4 &   $1.0\times 10^{-5}$ &  0.08 & 0.017 & 0.005 & 0.016 & 8.131 & 5577 & 1597  & 0.049 & 0.086 & $2.0\times 10^{-5}$ & {\it c} & 0.38\\
D5 &   $1.0\times 10^{-5}$ &  0.16 & 0.026 & 0.005 & 0.025 & 8.045 & 8998 & 1444  & 0.057 & 0.043 & $1.0\times 10^{-5}$ & {\it c} & 0.38\\
D6 &   $1.0\times 10^{-5}$ &  0.32 & 0.028 & 0.004 & 0.027 & 8.166 & 9529 & 1216  & 0.059 & 0.021 & $5.1\times 10^{-6}$  & {\it c}+{\it ac} & 0.44+0.44\\
D7 &   $1.0\times 10^{-5}$ &  0.64 & 0.013 & 0.003 & 0.012 & 8.103 & 4360 & 904   & 0.057 & 0.011 & $2.5\times 10^{-6}$ & {\it ac} & 0.38\\
\enddata
\end{deluxetable*}

\begin{figure*}[!htbp]
\centering
\includegraphics[width=\textwidth]{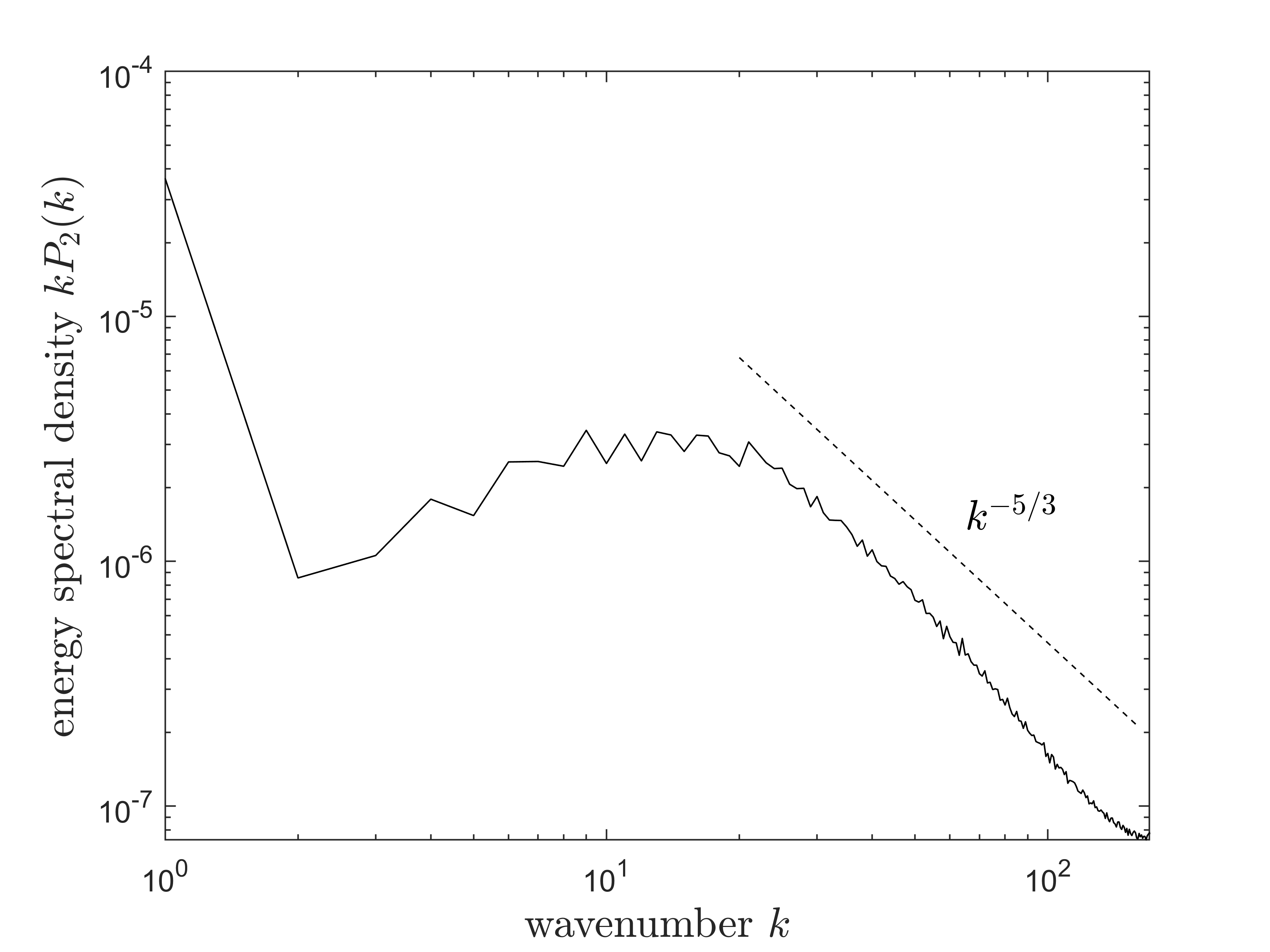}
\caption{Time averaged compensated power spectral density of the kinetic energy $kP_{2}(k)$ as a function of wavenumber $k$ at $z=0.5$ for case D7b. The averaged period is about 14 system rotation periods. The Kolmogrov $k^{-5/3}$ scaling is shown with the dashed line for reference.\label{fig:f6}}
\end{figure*}

To investigate whether LSVs could be formed at low latitudes, we performed a total of 9 simulations with different fluxes and angular velocities in the tilted {\it f}-plane at the latitude $\theta=22.5^{\circ}$. The basic settings of B5b-B7b, C5b-C7b, and D5b-D7b are all the same to the simulation Cases B5-B7, C5-C7, and D5-D7 at the latitude $\theta=90^{\circ}$. The only difference is that the angular velocity has two components in the tilted {\it f}-plane: the vertical component $\Omega \sin\theta$ along the $z$-direction, and the horizontal component $\Omega\cos\theta$ along the $y$-direction. The symbol `{\it s}' in the last column denotes shear flow. The detailed parameters of each simulation case are given in Table~\ref{table:tab2}. Other parameters and symbols have the same meanings as mentioned in Table~\ref{table:tab1}. At a given $z$, the velocity $\boldsymbol{v}$ can be decomposed into the summation of a Fourier series of $\boldsymbol{v}_{mn}$.  Then the two-dimensional kinetic energy spectrum \citep{chan1996turbulent,cai2018numerical} can be evaluated as
\begin{eqnarray}
\int P_{2}(k)dk=\sum_{m}\sum_{n}(v_{x,mn}v_{x,mn}^{*}+v_{y,mn}v_{y,mn}^{*}+v_{z,mn}v_{z,mn}^{*})~,
\end{eqnarray}
where the symbol star denotes the conjugate of a variable. Fig.~\ref{fig:f6} shows the time averaged compensated power spectral density of the kinetic energy $kP_{2}(k)$ as a function of $k$ for case D7b. The simulation marginally shows the Kolmogrov inertial range.

\startlongtable
\begin{deluxetable*}{cccccccccccccc}
\tablecaption{Parameters of simulation cases of {\it f}-plane at low latitude $\theta=22.5^{\circ}$\label{table:tab2}}
\tablehead{
 Case & $F_{tot}$ & $\Omega$ & $\langle v''\rangle$ & $\langle v_{z}''\rangle$ & $\langle v_{h}''\rangle$ & $Pr$ & $ Re$ & $Re_{z}$  & $ Ma$ & $ Ro_{c}$  & $E$ & {LSV}
}
\startdata
B5b &   $1.0\times 10^{-3}$ &  0.64 & 0.086 & 0.025 & 0.082 & 0.343 & 6905  & 1945 & 0.275 & 0.052 & $1.0\times 10^{-5}$ & {\it c} \\
B6b &   $1.0\times 10^{-3}$ &  1.28 & 0.113 & 0.023 & 0.110 & 0.335 & 9491  & 1844 & 0.342 & 0.026 & $5.0\times 10^{-6}$ & {\it c}+{\it ac} \\
B7b &   $1.0\times 10^{-3}$ &  2.56 & 0.111 & 0.021 & 0.109 & 0.336 & 9439  & 1712 & 0.375 & 0.013 & $2.5\times 10^{-6}$ & {\it c}+{\it ac} \\
\hline
C5b &   $1.0\times 10^{-4}$ &  0.32 & 0.030 & 0.012 & 0.028 & 1.672 & 4880  & 1870 & 0.118 & 0.047 & $1.0\times 10^{-6}$ & {\it s} \\
C6b &   $1.0\times 10^{-4}$ &  0.64 & 0.046 & 0.011 & 0.045 & 1.663 & 7753  & 1743 & 0.156 & 0.023 & $5.0\times 10^{-6}$ & {\it c}+{\it ac} \\
C7b &   $1.0\times 10^{-4}$ &  1.28 & 0.045 & 0.089 & 0.044 & 1.568 & 8810  & 1530 & 0.173 & 0.012 & $2.4\times 10^{-6}$ & {\it c}+{\it ac} \\
\hline
D5b &   $1.0\times 10^{-5}$ &  0.16 & 0.017 & 0.005 & 0.017 & 7.984 & 5939  & 1726 & 0.059 & 0.043 & $9.8\times 10^{-6}$ & {\it c} \\
D6b &   $1.0\times 10^{-5}$ &  0.32 & 0.023 & 0.004 & 0.023 & 7.765 & 8398  & 1499 & 0.066 & 0.022 & $4.8\times 10^{-6}$  & {\it c}+{\it ac} \\
D7b &   $1.0\times 10^{-5}$ &  0.64 & 0.015 & 0.004 & 0.014 & 7.657 & 5317  & 1335 & 0.061 & 0.011 & $2.4\times 10^{-6}$ & {\it c}+{\it ac}  \\
\enddata
%\tablecomments{Parameters of simulation cases of {\it f}-plane at low latitude $\theta=22.5^{\circ}$}
\end{deluxetable*}

\section{Movies}\label{appendixE}
\begin{figure}[!htbp]
\begin{interactive}{animation}{MovieS1.mpg}
\plotone{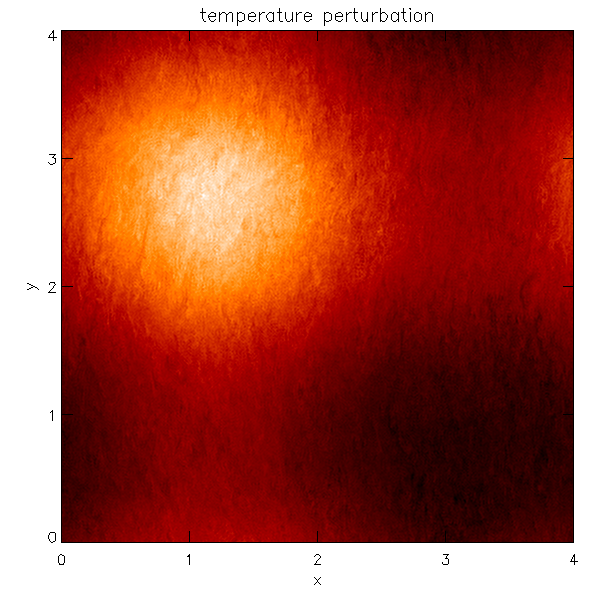}
\end{interactive}
\caption{Time evolution of temperature perturbation at $z=0.1$ of Case D7b. The time period is about 27 system rotation periods. The bright (dark) color denotes higher (lower) temperature.}\label{fig:movie1}
\end{figure}

\begin{figure}[!htbp]
\begin{interactive}{animation}{MovieS1.mpg}
\plotone{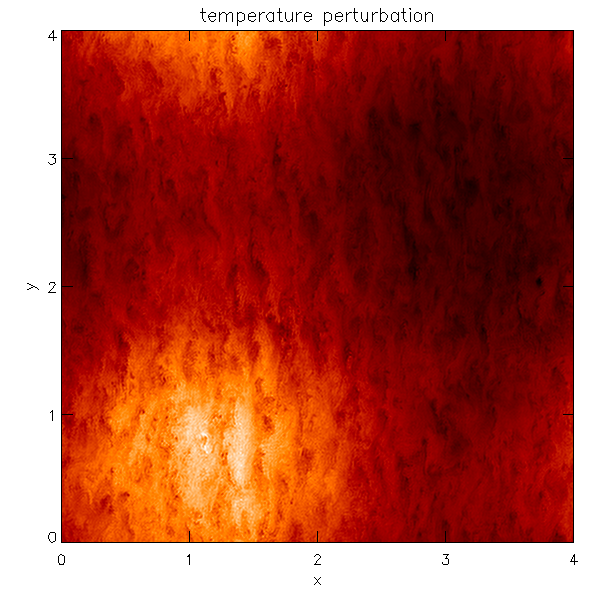}
\end{interactive}
\caption{Time evolution of temperature perturbation at $z=0.9$ of Case D7b. The time period is about 27 system rotation periods. The bright (dark) color denotes higher (lower) temperature.}\label{fig:movie2}
\end{figure}

\begin{figure}[!htbp]
\begin{interactive}{animation}{MovieS1.mpg}
\plotone{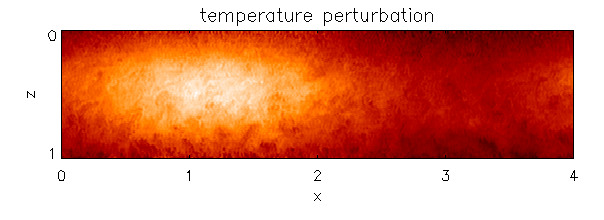}
\end{interactive}
\caption{Time evolution of temperature perturbation at $y=3.6$ of Case D7b. The time period is about 27 system rotation periods. The bright (dark) color denotes higher (lower) temperature.}\label{fig:movie3}
\end{figure}

\begin{figure}[!htbp]
\begin{interactive}{animation}{MovieS1.mpg}
\plotone{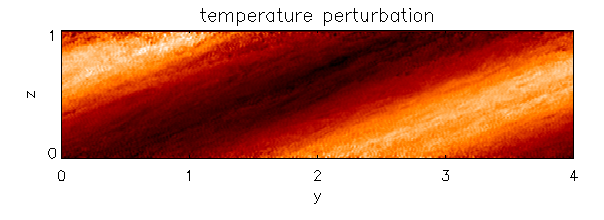}
\end{interactive}
\caption{Time evolution of temperature perturbation at $x=1.0$ of Case D7b. The time period is about 27 system rotation periods. The bright (dark) color denotes higher (lower) temperature.}\label{fig:movie4}
\end{figure}

\begin{figure}[!htbp]
\begin{interactive}{animation}{MovieS1.mpg}
\plotone{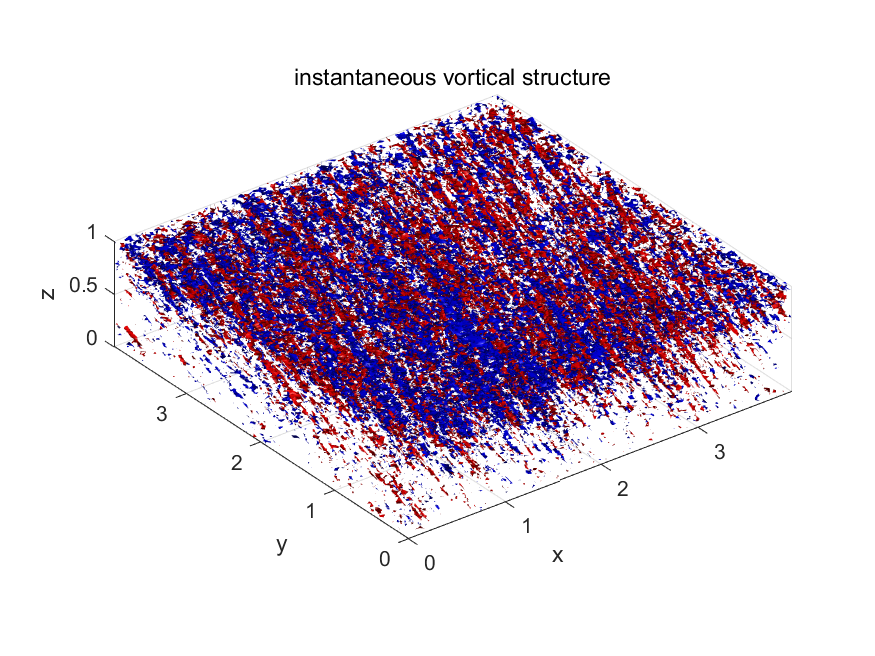}
\end{interactive}
\caption{Time evolution of vortical structure with positive (red color) and negative (blue color) vertical vorticity of Case D7b. The time period is about 27 system rotation periods.}\label{fig:movie5}
\end{figure}

The animations Figures \ref{fig:movie1}-\ref{fig:movie4} show the time evolutions of temperature perturbations at $z=0.1$, $z=0.9$, $y=3.6$, $x=1.0$ of Case D7b., respectively. The animations Figure \ref{fig:movie5} show the time evolution of vortical structure with positive (red color) and negative (blue color) vertical vorticity of Case D7b. The time period for each movie is about 27 system rotation periods.

%\bibliographystyle{aasjournal}
\bibliography{main}

\end{document}